\documentclass[article]{biometrika}
\usepackage{etex}

\usepackage{amsmath}
\usepackage{amssymb,amsfonts,latexsym,graphicx}
\usepackage{times}
\usepackage{wasysym,MnSymbol}%
\usepackage{bm}
\usepackage{natbib}
\usepackage[plain,noend]{algorithm2e}

\newcommand{\ind}{\Large\mbox{$\perp \kern-5.5pt \perp$}
 \newcommand{\nind}{\mbox{$\not\hsp\Delta{-4pt}\ind$}}}

\makeatletter
\renewcommand{\algocf@captiontext}[2]{#1\algocf@typo. \AlCapFnt{}#2} 
\def\@algocf@capt@plain{top}
\renewcommand{\algocf@makecaption}[2]{%
  \addtolength{\hsize}{\algomargin}%
  \sbox\@tempboxa{\algocf@captiontext{#1}{#2}}%
  \ifdim\wd\@tempboxa >\hsize
    \hskip .5\algomargin%
    \parbox[t]{\hsize}{\algocf@captiontext{#1}{#2}}
  \else%
    \global\@minipagefalse%
    \hbox to\hsize{\box\@tempboxa}
  \fi%
  \addtolength{\hsize}{-\algomargin}%
}
\makeatother

\usepackage{accents}
\newcommand{\ubar}[1]{\underaccent{\bar}{#1}}
\usepackage{booktabs}
\usepackage{amsmath}
\newcommand\numberthis{\addtocounter{equation}{1}\tag{\theequation}}

\usepackage{subcaption}
\usepackage{thmtools}
\declaretheoremstyle[%
spaceabove=3pt,%
spacebelow=3pt,%
headfont=\normalfont\rmfamily,%
postheadspace=1em,%
qed=,%
headpunct={}
]{mystyle} \declaretheorem[name={},style=mystyle,numberwithin=]{assumptionn}

\usepackage{tikz}
\usetikzlibrary{arrows,shapes.arrows,shapes.geometric,
	shapes.multipart,backgrounds,decorations.pathmorphing,positioning,fit,automata}

\usepackage{enumitem}
\usepackage{dcolumn}
\newcolumntype{d}[1]{D{.}{\cdot}{#1} }
\newcommand{\tx}{docetaxel plus
	estramustine }
\newcommand{\ctrl}{mitoxantrone plus prednisone}

\usepackage[none]{hyphenat}

\begin{document}

\jyear{}
\jvol{}
\jnum{}


\markboth{L. Wang, X. Zhou and T. Richardson}{Causal Inference with Truncation by Death}

\title{Identification and Estimation of Causal Effects with Outcomes Truncated by Death}

\author{LINBO WANG}
\affil{Department of Biostatistics, Harvard School of Public Health, 677 Huntington Avenue, Boston,
	Massachusetts 02115, U.S.A.
 \email{linbowang@g.harvard.edu}}

\author{ XIAO-HUA ZHOU}
\affil{Department of Biostatistics, University of Washington, Seattle, Washington 98195, U.S.A.
\email{azhou@u.washington.edu}}

\author{\and THOMAS S. RICHARDSON}
\affil{Department of Statistics, University of Washington, Box 354322, Washington 98195, U.S.A.
 \email{thomasr@u.washington.edu}}

\maketitle

\begin{abstract}
	It is common in medical studies that the outcome of interest is truncated by death, meaning that a subject has died before the outcome could be measured. In this case, restricted analysis among survivors may be subject to selection bias. Hence, it is of interest to estimate the survivor average causal effect, defined as the average causal effect among the subgroup consisting of subjects who would survive under either exposure.  In this paper, we consider the identification and estimation problems of the survivor average causal effect. We propose to use a substitution variable in place of the latent membership in the always-survivor group. The identification conditions required for a substitution variable are conceptually similar  to conditions for a conditional instrumental variable, and may apply to  both randomized and observational studies. We show that the survivor average causal effect is identifiable with use of such a substitution variable, and propose novel model parameterizations for estimation of the survivor average causal effect under our identification assumptions. Our approaches are illustrated via simulation studies and a data analysis.
\end{abstract}


\section{Introduction}

%
%
%

In medical studies, researchers are often interested in evaluating risk factors for a non-mortality outcome. However, this outcome may be truncated by death and hence, undefined if some subjects  die before the follow-up assessment.  For example, suppose we are interested in estimating the effect of smoking on memory decline in an aged population. If a subject dies before the follow-up memory test is administered, then his/her memory score at the follow-up visit is undefined. Direct comparisons between smokers and non-smokers among observed survivors are subject to selection bias, as non-smokers are more likely to survive to the follow-up assessment, and survival is associated with memory decline  \citep{rosenbaum1984consquences, robins1992identifiability}.
More fundamentally,  direct comparisons among observed survivors  are not causally interpretable, as they compare outcomes from different subpopulations at baseline \citep{rubin2006causal}. \citet[\S 12.2]{robins1986new} proposed to estimate the average causal effect in the always-survivor group,
the group of subjects who would survive if they chose to receive either exposure at baseline.  The always-survivor group was later termed a {principal stratum} by  \cite{rubin1998more} and \cite{frangakis1999addressing,frangakis2002principal},
and the  contrast within this group is the survivor average causal effect. This contrast
 is causally interpretable because membership of the always-survivor group is defined at baseline, and since these subjects always survive, their subsequent outcomes under both treatments are well-defined.   Alternative estimands 
are  discussed in \cite{kurland2009longitudinal} and \cite{weuve2012accounting}.

The survivor average causal effect is not identifiable without further assumptions \citep{zhang2003estimation}. Large sample bounds for the survivor average causal effect have been derived under minimal assumptions \citep{zhang2003estimation,imai2008sharp,long2013sharpening}.  In order to identify this effect, it is common  to perform a sensitivity analysis  by assuming a class of identification conditions indexed by a sensitivity parameter \citep{gilbert2003sensitivity, shepherd2006sensitivity, hayden2005estimator, egleston2007causal,jemiai2007semiparametric,chiba2011simple}.  

Alternatively, identification of the survivor average causal effect can be based on covariate information. For example, \cite{tchetgen2014identification} introduced a {variant} of the survivor average causal effect that is identified when risk factors of survival are available in post-exposure follow-ups. The resulting causal contrast is for a principal stratum defined in term of particular post-exposure risk factors.
Hence, in principle, the causal estimand can differ depending on the set of post-exposure risk factors used in the analysis.
\cite{ding2011identifiability} took a different approach to identifying the survivor average causal effect in a randomized study setting. They	  proposed a semiparametric identification method based on 
a baseline variable whose distribution is informative of the membership of the always-survivor group. Using this baseline variable, they showed that the survivor average causal effect was identifiable under their assumptions.	However, as pointed out by \cite{tchetgen2014identification},  their assumption essentially requires that there are no common causes of the survival and outcome processes, which is very unlikely even in randomized studies.

In this article, we relax \cite{ding2011identifiability}'s identification assumptions by employing more detailed covariate information. In contrast to \cite{tchetgen2014identification}'s estimand, our causal parameter is  defined independently of the covariates incorporated in the analysis. Furthermore, the proposed approach is applicable to both randomized trials and observational studies,  {and} allows for {measured} common causes of the survival and outcome processes. We also discuss possible violations and alternative approaches to our identification model. 
To estimate the survivor average causal effect in practice, we impose additional distributional assumptions. This is challenging for two reasons. First, unlike the standard observational study setting, in our identification framework, the baseline covariates may be common causes between any pair of treatment, survival and outcome. As we explain in Section \ref{sec:est}, this unique role for baseline covariates makes the standard propensity score methods inappropriate in our setting. Second, novel model parameterizations are needed to make our distributional assumptions compatible with our  identification assumptions.

\section{Data structure, notation and causal estimand}
\label{sec:notations}

Consider a medical study with a single follow-up visit.
Let $Z$ be the exposure indicator and  $W$ denote covariates observed at baseline.  We assume that each subject has two potential survival outcomes $S(1)$ and $S(0)$, defined as the survival statuses at the follow-up visit that would have been observed had the subject been exposed and unexposed, respectively.  Similarly we let $Y(1)$ and $Y(0)$ denote the potential non-mortality outcomes under exposure and non-exposure, respectively. We assume that for $z=0,1$, $Y(z)$ takes real values only  if $S(z)=1$.  We extend the definition of $Y(z)$ so that it takes  the constant value $*$ if $S(z)=0$. 

We use $G$ to denote the
survival type as defined in Table
\ref{table:defG}, in which L denotes live and D denotes die. One can see from Table \ref{table:defG} that  there exists a one-to-one mapping between the  survival type and the bivariate potential survival $\{S(1),S(0)\}$, so $G$ can be interpreted as an abbreviation for $\{S(1),S(0)\}$. 

We adopt the axiom of consistency such that  the observed outcome $Y$ satisfies $
Y = Z Y(1) + (1-Z) Y(0)
$ and the observed survival $S$ satisfies $S = Z S(1) + (1-Z) S(0)$.
The observed samples  $O_i = (Z_i, W_i, S_i, Y_i), i=1,\ldots,N$ are independently drawn from an infinite super-population.

\begin{table}[!htbp] 
	\centering
	\caption{Patient survival types  }\label{table:defG}
	\begin{tabular}{ccccc}
		$S(1)$ & $S(0)$ & Survival type & $G$  &  Description \\[5pt]
		1 & 1 &  always-survivor & \textsc{LL} & The subject always survives, regardless of exposure status. \\
		1 & 0 &  protected  & \textsc{LD}  & The subject survives if exposed, but dies if not exposed. \\
		0 & 1 &  harmed & \textsc{DL}  & The subject dies if exposed, but survives if not exposed. \\
		0 & 0 &  doomed & \textsc{DD}  &  The subject always dies, regardless of exposure status. \\
	\end{tabular}
\end{table}

Throughout this article, we assume that there is no interference between study subjects regarding both the survival $S$ and the non-mortality outcome $Y$, and there is only one version of exposure
\citep{rubin1980comment}.

\cite{rubin2000causal} noted that the observed survivors in the exposed group are from a mixture of always-survivor and protected strata, while the observed survivors in the non-exposed group are from a mixture of always-survivor and harmed strata. As a result,  direct comparisons between  different exposure groups among observed survivors are not causally meaningful as these people are from different subpopulations at baseline.  
To address this, since the always-survivor stratum is the only group for which $Y(1)$ and $Y(0)$ both take real values, we define our causal estimand to be  the average causal
effect in this stratum:
\begin{equation*}
\label{def:sace}
\Delta_{\textsc{LL}}  = E\{Y(1)-Y(0)\mid G=\textsc{LL}\} .
\end{equation*}
This estimand is also known as the survivor average causal effect.

\section{Identification of the survivor average causal effect}
\label{sec:identify}

\subsection{The identification problem}
\label{sec:assump}

Though causally interpretable, the survivor average causal effect is in general not identifiable as it depends on the potential outcomes $Y(1), Y(0), S(1)$ and $S(0)$. Furthermore, it is not identifiable even under assumptions that are sufficient for identification in other causal contexts:

%
%
%

\begin{assumptionn} [Monotonicity]
	\label{assump.monotone} {$S(1) \geq S(0)$ almost surely}.
\end{assumptionn}

The monotonicity assumption may be plausible in some observational studies. For example, in studies evaluating the effect of smoking on memory decline, it is widely believed that smoking is bad for overall health, and hence overall survival. This assumption tends to be questionable in randomized clinical trials with acute diseases because typically a clinical trial would be unethical if the researchers believe that  one treatment benefits survival a priori. To address this issue, we relax this assumption later in Section \ref{sec:relax_momotone}.

\begin{assumptionn}[S-Ignorability]
	\label{assump:s_igno}
	The treatment assignment is independent of the potential survivals given the observed covariates $W$, so that
	$
	Z  \ind  S(z)\mid W; z=0,1.
	$
\end{assumptionn}	

\begin{assumptionn}[Y-Ignorability] 
	\label{assump.ll.igno}
	The treatment assignment in the always-survivor stratum is independent of the potential outcomes given observed covariates $W$, so that $Z \ind Y(z)\mid W,G=\textsc{ll}; z=0,1.$
\end{assumptionn}
\ref{assump:s_igno} and \ref{assump.ll.igno} are similar to the weakly ignorable treatment assignment assumption \citep[see e.g.,][]{imbens2000role}.
Under \ref{assump.ll.igno}, we have
\begin{flalign*}
E\{Y(z)\mid G=\textsc{LL}\} &=   \dfrac{E_W \left(\mu_{z,\textsc{LL},W} \pi_{\textsc{LL}\mid W}\right)}{E_W(\pi_{\textsc{LL}\mid W})},
\numberthis \label{eqn:sace}
\end{flalign*}
where $\mu_{z,g,W} = E(Y\mid Z=z,G=g,W)$ and $\pi_{g\mid W} = {\rm pr}(G=g\mid W)$. 
Under \ref{assump.monotone} and \ref{assump:s_igno}, $Z \ind (S(1),S(0))\mid W$,
so the $W$-specific survival-type probabilities are identified by
\begin{equation}
\label{eqn:pillW}  
	\pi_{\textsc{LL}\mid W} = {\rm pr}(S=1\mid Z=0,W), \ \ 
	\pi_{\textsc{LD}\mid W} = {\rm pr}(S=1\mid Z=1,W) - {\rm pr}(S=1\mid Z=0,W);
\end{equation}
$\mu_{0,\textsc{LL},W}$ can be identified by
\begin{flalign}
\label{eqn:mu0}
\mu_{0,\textsc{LL},W} = E(Y\mid Z=0,S=1,W).
\end{flalign}
However, $\mu_{1,\textsc{LL},W}$ is not identifiable from the observed data. In fact, as noted by \cite{zhang2003estimation}, 
the observed data in the group $\{i: Z_i = 1, S_i=1, W_i = w_i\}$ can be written as a mixture of two distributions from the $\textsc{LL}$ and $\textsc{LD}$ strata:
$$
	F(y\mid Z=1, S=1, W) = \sum\limits_{g=\textsc{LL},\textsc{LD}}p_{g\mid 1, W, s=1} F(y\mid Z=1, G=g, W),
$$
where $F$ denotes the cumulative distribution function, and for $g=\textsc{LL}$ or $\textsc{LD}$, $p_{g\mid z,W, s=1} \equiv {\rm pr}(G=g\mid W) / {\rm pr}(S=1\mid Z=z, W)$ is identifiable from data.  Unless the mixing probability $p_{\textsc{LL}\mid 1,w}$ is $1$ or $0$, $\mu_{1,\textsc{LL},W}$ is only partially identifiable in that  there is a range of values for $\mu_{1,\textsc{LL},W}$ that are compatible with the observed data distribution.

\subsection{Identifying the survivor average causal effect using a substitution variable}
\label{sec:iden}

To identify the survivor average causal effect, without loss of generality, we assume that the baseline covariates $W$ can be written as $(X,A)$. We propose an identification framework in which the role of $X$ is similar to a confounder whereas that of $A$ is similar to an instrument.
Specifically, we make the following assumptions on $A$ and $X$:

\begin{assumptionn}[Exclusion restriction]
	\label{assump.A2.x} 
	$
	A \ind Y(1)\mid  Z=1, G, X.
	$
\end{assumptionn}

\begin{assumptionn}[Substitution relevance]
	\label{assump.A3.x}
	$A \not\ind G\mid Z=1,S=1,X.$
\end{assumptionn}

The survival type $G$ is a latent baseline variable that satisfies \ref{assump.A2.x} and \ref{assump.A3.x}. For this reason, any variable $A$ satisfying \ref{assump.A2.x} and \ref{assump.A3.x} is called a {substitution variable} for $G$.
The conditions for a substitution variable are similar to those for an instrumental variable \citep{angrist1996identification}. Specifically, \ref{assump.A2.x} is similar to the exclusion restriction assumption  in an instrumental variable analysis in that they both capture the notion of $A$ having no direct effect on $Y$, and \ref{assump.A3.x} is similar to the instrumental relevance  assumption; in particular, they both require that $A$ be a relevant variable. {We clarify that similar to the instrumental relevance assumption, \ref{assump.A3.x} needs to hold for all possible values of $X$.} As we illustrate in Section \ref{sec:illustration}, even in a randomized study setting, the inclusion of covariate information $X$ makes \ref{assump.A2.x} and \ref{assump.A3.x}  more plausible.

Theorem \ref{thm:identification} below states that the survivor average causal effect is identifiable with a substitution variable for the survival type.  Proofs of theorems are left to the Supplementary Material.

\begin{theorem}
	\label{thm:identification}
	Under \ref{assump.monotone}--\ref{assump.A3.x}, $\Delta_{\textsc{LL}}$  is identifiable.

\end{theorem}

When $A$ takes more than two values, $\Delta_{\textsc{LL}}$  may be over-identified. In this case,  one may falsify our identification assumptions using specification tests such as  the Sargan--Hansen test.

\subsection{A nonparametric structural equation model with independent errors}
\label{sec:illustration}

\begin{figure}[!htbp] 
	\centering
\scalebox{0.8}{
	\begin{tikzpicture}[->,>=stealth',shorten >=1pt,auto,node distance=2.5cm, shape=ellipse,
	semithick, scale=0.3]
	pre/.style={->,>=stealth,semithick,blue,line width = 1pt}]
	\tikzstyle{every state}=[fill=none,draw=black,text=black, shape=ellipse]
	\node[state] (X)                    {$X$};
	\node[state] (A) [right of=X] {$A$};
	\node[state]         (Z) [right of=A] {$Z$};
	\node[state](S) [right of = Z]{$S$};
	\node[state](Y)[right of = S]{$Y$};
	\path	(X) edge node {} (A)
	(A) edge node {} (Z)
	(Z) edge node {} (S)
	(S) edge node  {} (Y);
	\draw[->] (X)  to[out=20,in=160,looseness=1.1] (Z);
		\draw[->] (X) to[out=40,in=160,looseness=0.8] (S);
	\draw[->] (X) to[out=-20,in=-140,looseness=0.9] (Y);
		\draw[->, bend right] (A) edge (S);
		\draw[->] (Z) to[out=-20,in=-160,looseness=1.1] (Y);
	\end{tikzpicture}	
 }
\caption{The simplest causal diagram associated with the structural equations \eqref{eqn:npsem}.}
\label{fig:npsem}
\end{figure}
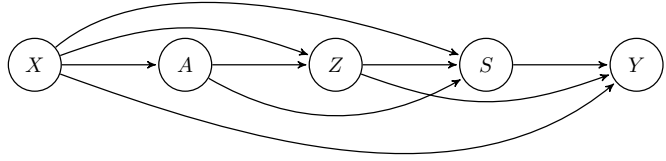

The key assumptions in our identification model are  \ref{assump:s_igno}--\ref{assump.A2.x}. 
These  are implied by a certain nonparametric structural equation model with independent errors  \citep{pearl2009causality}  given by 
\begin{equation}
\label{eqn:npsem}  
X = f_X(\epsilon_X), A  = f_A(X, \epsilon_A),  Z  = f_Z(X, A, \epsilon_Z), 
S = f_S(X,A,Z,\epsilon_S), 
Y = f_Y(X,Z,S,\epsilon_Y), 
\end{equation}
where the error terms  $\epsilon_X, \epsilon_A, \epsilon_Z, \epsilon_S, \epsilon_Y$  are jointly independent.  
Here the error terms can be interpreted as the set of one-step-ahead counterfactuals; for example, $\epsilon_Z$ can be interpreted as $\{ Z(a,x): a\in \mathcal{A}, x \in \mathcal{X}\}$, where $\mathcal{A}$ and $\mathcal{X}$ are the sample space for $A$ and $X$, respectively.  
Figure \ref{fig:npsem} gives the simplest causal diagram associated with  \eqref{eqn:npsem}. There is no directed edge from $A$ to $Y$, encoding the notion that $A$ has no direct effect on $Y$. We remark that  \ref{assump:s_igno}--\ref{assump.A2.x} also follow from \eqref{eqn:npsem} with dependent errors that allow for unmeasured confounding between nodes in Fig. \ref{fig:npsem}. See the Supplementary Material for examples.

In contrast to the identification assumptions in \cite{ding2011identifiability}, we include baseline covariates $X$ in our identification framework. Our approach is  applicable to observational studies since we allow for the edge $X \rightarrow Z$. Moreover, we allow for edges $X \rightarrow S$ and $X\rightarrow Y$, thereby avoiding the assumption that there are no common causes of the survival and outcome processes, which is very restrictive even for randomized studies.

\begin{remark}
The assumptions of a nonparametric structural equation model with independent errors   cannot be tested, even in principle, via any randomized experiment on the measured variables \citep{robins2010alternative}.	
\end{remark}

\subsection{{Alternatives to the individual-level monotonicity assumption}}
\label{sec:relax_momotone}

We now consider alternative stochastic monotonicity assumptions  in place of the individual-level assumption \ref{assump.monotone}. These alternatives were used  by \cite{roy2008principal} in the context of non-compliance  and by \cite{lee2010causal}  in the context of truncation by death.

We characterize the association between the potential survivals $S(1)$ and $S(0)$ by
$$
\rho(W)  = \dfrac{{\rm pr}(G=\textsc{LL}\mid W) - {\rm pr}\{S(1)=1\mid W\} {\rm pr}\{S(0)=1\mid W\} }{\min[{\rm pr}\{S(1)=1\mid W\},{\rm pr}\{S(0)=1\mid W\} ] - {\rm pr}\{S(1)=1\mid W\} {\rm pr}\{S(0)=1\mid W\} }.
$$
Note that $\rho(W)$ is no larger than 1 and  differs from the correlation between $S(1)$ and $S(0)$ by a non-negative identifiable factor. In particular, the sign of $\rho(W)$ coincides with the sign of the conditional correlation cor$\{S(1), S(0) \mid W\}$.  In practice, one might be willing to assume that the potential survivals under active treatment and control are not negatively correlated such that 
\begin{equation*}
\label{eqn:restriction}
{\rm pr}\{S(1)=1\mid S(0)=1,W\} \geq {\rm pr}\{S(1)=1\mid S(0)=0, W\}  
\end{equation*}
almost surely. This gives rise to the following stochastic monotonicity assumption:
\begin{assumptionn}[Stochastic monotonicity]
	\label{assump:stochastic_monotone}
	$$0 \leq \rho(W) \leq 1 $$ 
 almost surely. 
\end{assumptionn}

Note that $\rho(W) = 0$ if and only if  conditional on observed covariates $W$, $S(1)$ is not correlated with $S(0)$. On the other hand, $\rho(W) = 1$ if and only if ${\rm pr}(G=\textsc{LL}\mid W) = \min\left[{\rm pr}\{S(1)=1\mid W\},{\rm pr}\{S(0)=1\mid W\} \right]$.
We emphasize that \ref{assump:stochastic_monotone} does not reduce to \ref{assump.monotone} when $\rho(W) = 1$; in  particular,
assuming  a value for $\rho(W) \in [0,1]$ does not specify which treatment is more beneficial for survival. 
This is particularly appealing when the monotonicity assumption is not plausible {a priori} due to ethical reasons.

Theorem \ref{thm:identification_alter1} states that under \ref{assump:s_igno}--\ref{assump:stochastic_monotone}, if we further assume that \ref{assump.A2.x} and \ref{assump.A3.x} hold in the control arm, then given $\rho(W)$, the survivor average causal effect is identifiable from the observed data. The additional conditional independence condition (i) in Theorem \ref{thm:identification_alter1} also follows from the structural equation model \eqref{eqn:npsem} with independent errors. 

\begin{theorem}
	\label{thm:identification_alter1}
	If we assume \ref{assump:s_igno}--\ref{assump:stochastic_monotone}, and  that 
	\begin{enumerate}
		\item [(i)] $A \ind Y(0)\mid Z=0,G,X$;
		\item [(ii)] for all $x\in \mathcal{X}$,
		$A \not\ind G\mid Z=0,S=1,X=x$,
	\end{enumerate}
	then $\Delta_{\textsc{LL}}$  is  identifiable given $\rho(W)$.
\end{theorem}

\subsection{Relaxing the exclusion restriction assumption}

In practice, the ignorability assumptions \ref{assump:s_igno} and \ref{assump.ll.igno} may be questionable if there are {unmeasured} confounders between the exposure $Z$ and either the survival $S$ or the non-mortality outcome $Y$. Similarly, 
the exclusion restriction assumption \ref{assump.A2.x} may be questionable {if the set of observed covariates $X$ is not rich enough to contain all common causes of
the non-mortality  outcome 
and any one of the treatment, the substitution variable and survival.}
There has been extensive literature on relaxing the ignorability assumptions for causal inference in observational studies \citep[see e.g.,][]{rosenbaum1983assessing,imbens2003sensitivity,ding2014generalized}, which may be applied here with slight modification. Hence in this article, we focus on alternative assumptions to relax the exclusion restriction assumption.

%
%

%

Specifically, we propose an alternative  no-interaction assumption \ref{assump.A2.x.alter}, which assumes that conditioning on the baseline variables $X$, neither $Z$ nor $G$ modifies the effect of $A$ on $Y$. 

\begin{assumptionn} (no-interaction)
	\label{assump.A2.x.alter} 
	For all $x, a_0, a_1$, 
	\begin{flalign}
	&\quad    E(Y\mid Z=1,G=\textsc{LD},X=x,A=a_1) -     E(Y\mid Z=1,G=\textsc{LD},X=x,A=a_0) \label{eqn:a71} \\  &=     E(Y\mid Z=1,G=\textsc{LL},X=x,A=a_1) -     E(Y\mid Z=1,G=\textsc{LL},X=x,A=a_0)  \label{eqn:a72}   \\ 
	&= E(Y\mid Z=0,G=\textsc{LL},X=x,A=a_1) - E(Y\mid Z=0,G=\textsc{LL},X=x,A=a_0). \label{eqn:a73}
	\end{flalign}
\end{assumptionn}
In contrast, the exclusion restriction assumption \ref{assump.A2.x} implies that the  comparisons in \eqref{eqn:a71} and \eqref{eqn:a72} are both 0. 

 This no-interaction assumption may be plausible when the substitution variable affects the outcome independently of other disease processes. For example, suppose one is interested in studying the effects of statin usage on memory decline, taking baseline dementia status as the substitution variable. It is likely that baseline dementia status will affect the outcome directly as memory decline is an irreversible process: demented subjects are more likely to experience decline in memory. However, it seems reasonable to assume that neither statin usage, which is typically due to high blood cholesterol levels, nor the latent survival statuses, which largely reflect baseline cholesterol levels, would interact, on the additive scale, with the process of memory decline. 

To relax \ref{assump.A2.x} in the absence of $X$,  \cite{ding2011identifiability}  also assumed that \eqref{eqn:a71} is equal to \eqref{eqn:a72}.  In addition, they placed linearity assumptions on $E(Y\mid Z=1,G,A)$, and  required that the substitution variable $A$ had at least three categories or was continuous.
{Both assumptions may be overly restrictive in practice. For example, they may not hold in the example of statin usage and memory decline described above,
taking $A$ to be dementia status at baseline. This is because current knowledge on dementia suggests that  baseline cognitive status  has a non-linear influence on the deterioration of
cognitive score  \citep[e.g.,][]{sperling2014evolution}; in addition,
baseline dementia status is often coded as a binary variable.}
 To mitigate the need for linearity identification assumptions and allow for binary substitution variables, we instead assume that  \eqref{eqn:a72} is equal to \eqref{eqn:a73}; that is, $Z$ is not an effect modifier for the conditional effect of $A$ on $Y$.  This is similar in spirit to 
the no-interaction assumption in causal mediation analysis \citep[e.g.,][]{robins2003semantics}.  In particular, it is guaranteed to hold  under the strong null of no treatment effect of $Z$ on $Y$. 

The inclusion of baseline covariates in \ref{assump.A2.x.alter} has important implications for the {design} of studies with truncation by death:
every possible cause of the outcome should be measured, if feasible, to try to ensure that 
 no residual effect modification remains within strata defined by $X$.

Theorem \ref{thm:identification_alter} states  that if we replace the exclusion restriction assumption  \ref{assump.A2.x} with  \ref{assump.A2.x.alter}, then the survivor average causal effect is still identifiable.  
\begin{theorem}
	\label{thm:identification_alter}
	Under assumptions \ref{assump.monotone}--\ref{assump.ll.igno}, \ref{assump.A3.x} and \ref{assump.A2.x.alter}, $\Delta_{\textsc{LL}}$  is  identifiable.
	\end{theorem}

\section{Model parameterization}
\label{sec:est}

In the previous section we have shown that the survivor average causal effect is identifiable under various identification assumptions. {When the covariates $W$ take only a few discrete values, within each level of $X$, the identification formulae in the proofs of Theorem \ref{thm:identification}--\ref{thm:identification_alter}  imply plug-in estimators in the case of binary  $A$  and generalized method of moment estimators in the case that $A$ takes more than two discrete values   \citep{ding2011identifiability,hansen1982large}.} When $W$ are continuous or high dimensional, however, we need to impose additional distributional assumptions. We first note that the identification assumptions imply certain constraints on the observed data law, as described in Proposition \ref{prop:constraints}.

\begin{proposition}
	\label{prop:constraints}  \quad \\[-18pt]
	
	\begin{itemize}
		\item[(I)] The assumptions of Theorem \ref{thm:identification} imply the following constraints on the law of $(Z,S,Y,W)$:
		\begin{flalign}
		{\rm pr}(S=1\mid Z=0, X, A)  &\leq {\rm pr}(S=1\mid Z=1, X, A); \label{eqn:constraint_monotone}\\
		\text{for all } x,			E(Y\mid Z=1, S=1, X=x, A=a) &\text{ is bounded as a function of } a; \label{eqn:constriant_bounded} \\
			\text{for all } x,	\dfrac{{\rm pr}(S=1\mid Z=0,X=x,A=a)}{{\rm pr}(S=1\mid Z=1,X=x,A=a)} &\text{\ is not   constant as a  function of } a. \label{eqn:dependence}
		\end{flalign}
		\item[(II)] {The assumptions of Theorem \ref{thm:identification_alter1} imply  \eqref{eqn:constriant_bounded} and that 
		\begin{equation}
			\text{for all } x,			E(Y\mid Z=0, S=1, X=x, A=a) \text{ is bounded as a function of } a. \label{eqn:constriant_bounded3}
		\end{equation}}
		\item[(III)] The assumptions of Theorem \ref{thm:identification_alter} imply \eqref{eqn:constraint_monotone} and that 
		\begin{flalign*}
		& \quad \text{ for all } x, 		 E(Y\mid Z=1, S=1, X=x, A=a) - E(Y\mid Z=0, S=1,  X=x, A=a) \\  &\text{ is bounded as a function of } a.  \numberthis \label{eqn:constriant_bounded2}
		\end{flalign*}
	\end{itemize}
\end{proposition}

To ensure that the modeling assumptions are compatible with the constraints in Proposition \ref{prop:constraints}, we avoid imposing distributional assumptions on the observed data directly. Instead, we make  the following distributional assumptions on the law of $\{Z,W,S(1),S(0),Y(1),Y(0)\}$, which are compatible with the identification assumptions. For brevity here we only give models under the assumptions of Theorem \ref{thm:identification}. Parameterizations for estimating $\Delta_{\textsc{LL}}$  under alternative identification assumptions  which are similar in spirit are described in Appendix \ref{appendix:models_alter}.
\begin{enumerate}[label=M.{\arabic*}]
	\item \label{m1.1} $E\{Y(0)\mid Z=0,G=\textsc{LL},X,A\}$ is known up to a finite-dimensional parameter $\alpha_1$; that is, $E\{Y(0)\mid Z=0,G=\textsc{LL},X,A\} = m_{1}(X,A; \alpha_1)$, where $m_{1}(\cdot,\cdot; \alpha_1)$ is a known function and $\alpha_1$ is an unknown parameter. 
	Specifically, for the simulations and data example we consider
	\begin{equation}
	\label{eqn:m1.1}
	m_{1}(X,A; \alpha_1) = \alpha_{10} +  X^T \alpha_{11} + A \alpha_{12}.
	\end{equation}
	\item  \label{m1.2} $E\{Y(1)\mid Z=1,X,G,A\}$ is known up to a finite-dimensional parameter $\alpha_2$; that is, $E\{Y(1)\mid Z=1,X,G,A\} = m_{2}(X,G; \alpha_2)$, where $G$ takes values in $\{\textsc{LL}, \textsc{LD}\}$, $m_{2}(\cdot,\cdot; \alpha_2)$ is a known function and $\alpha_2$ is an unknown parameter.  Note that due to \ref{assump.A2.x}, $E\{Y(1)\mid Z=1,X,G,A\}$ does not depend on the value of $A$. 
	Specifically, for the simulations and data example we code $\textsc{LL}$ to be 1, $\textsc{LD}$ to be 0, and consider
	\begin{equation}
	\label{eqn:m1.2}
	m_{2}(X,G; \alpha_2) = \alpha_{20} +  X^T \alpha_{21} +  G \alpha_{23}.
	\end{equation}
	\item   \label{m2.1} ${\rm pr}\{S(1)=1\mid X,A\}$ is known up to a finite-dimensional parameter $\beta_1$; that is ${\rm pr}\{S(1)=1\mid X,A\} = \theta_1(X,A;\beta_1)$, where $\theta_1(\cdot,\cdot;\beta_1)$ is a known function and $\beta_1$ is an unknown parameter.
	Specifically, for the simulations and data example we consider
	\begin{equation}
	\label{eqn:m2}
	\theta_1(X,A;\beta) = \text{expit}(\beta_{10}+  X^T \beta_{11} + A \beta_{12}).
	\end{equation}
	\item   \label{m2.2} ${\rm pr}\{S(0)=1\mid X,A\} / {\rm pr}\{S(1)=1\mid X,A\}$ is known up to a finite-dimensional parameter $\gamma$; that is ${\rm pr}\{S(0)=1\mid X,A\} / {\rm pr}\{S(1)=1\mid X,A\} = \theta_{0/ 1}(X,A;\gamma)$, where $\theta_{0/ 1}(\cdot,\cdot;\gamma)$ is a known function and $\gamma$ is an unknown parameter.  Specifically, for the simulations and data example we consider
	\begin{equation}
	\label{eqn:m3}
	\theta_{0/1}(X,A;\gamma)  = \text{expit}(\gamma_0+ X^T \gamma_1 +  A \gamma_2).
	\end{equation}
\end{enumerate}
An alternative approach to \ref{m2.1} and \ref{m2.2}  is to impose distributional assumptions on  ${\rm pr}\{S(z)\mid X,A\},$ $z=0,1$ directly \citep[e.g.,][]{lee2010causal}. However, constraint \eqref{eqn:constraint_monotone} implies that ${\rm pr}\{S(0)=1\mid X,A\}$ resides in the range $[0, {\rm pr}\{S(1)=1\mid X,A\}]$. Hence the model parameters for ${\rm pr}\{S(1)=1\mid X,A\}$ live in a constrained space, making estimation and asymptotic analysis difficult. To avoid such constraints, we {reparameterize our models} as in  \ref{m2.1} and \ref{m2.2}.

To derive the maximum likelihood estimator,  we note that \ref{m1.1}--\ref{m2.2} correspond to the following modeling constraints on the observed data distribution:
\begin{flalign*}
{\rm pr}(S=1\mid Z=1,X,A) &= 	\theta_1(X,A;\beta_1) , \\
{\rm pr}(S=1\mid Z=0,X,A) &= 	\theta_1(X,A;\beta_1)  	\theta_{0/1}(X,A;\gamma),  \\
E(Y\mid Z=1,S=1,X,A) &= \theta_{0/1}(X,A;\gamma) m_{2}(X,1; \alpha_2) + \{1-\theta_{0/1}(X,A;\gamma)\}  m_{2}(X,0; \alpha_2), \\
E(Y\mid Z=0,S=1,X,A) &= m_{1}(X,A; \alpha_1) .
\end{flalign*}	
It is easy to see that these models are compatible with the testable implications  described in Proposition \ref{prop:constraints}. 
Ordinary least squares and maximum likelihood estimation may be used for the parameter estimation.
$\Delta_{\textsc{LL}}$  can then be estimated using \eqref{eqn:sace}, in which $\mu_{z,\textsc{LL},W}$ can be estimated using \eqref{eqn:m1.1} and \eqref{eqn:m1.2}, $\pi_{\textsc{LL}\mid W}$ can be estimated using the product of \eqref{eqn:m2} and \eqref{eqn:m3}, and the integration is taken with respect to the empirical distribution of $W$.  
Using standard M-estimation theory, one can show that the resulting estimate of $\Delta_{\textsc{LL}}$  is consistent and asymptotically normally distributed.

\begin{remark}
	An alternative popular approach to estimating causal effects is based on the propensity score \mbox{$e(X)$}, defined as the probability of assignment to  exposure conditioning on baseline covariates.  However, although the propensity score is sufficient for summarizing the effect of \mbox{$X$} on \mbox{$Z$} in the sense that
	$
	Z  \ind X \mid e(X),
	$	
	it is not sufficient for summarizing the effect of $X$ on $A$. 
	 Thus  the propensity score methods are  not directly applicable for estimating the survivor average causal effect under our identification assumptions. 
\end{remark}

%
%
%
%
%
%
%
%
%
%
%

\section{Simulation studies}
\label{sec:simulation}


\subsection{Settings  following \ref{assump.monotone}--\ref{assump.A3.x}}
\label{sec:simu_1}

We first consider simulation settings in which data were generated according to Fig. \ref{fig:npsem}.
Specifically the baseline covariates $X=(X_1,X_2,X_3)$ are a combination of discrete and continuous
variables: $X_1$ is a discrete variable taking values 1 or -1 with probability 1/2, 
and  $(X_2,X_3)$ follows a multivariate normal distribution $N(\mu,\Sigma)$, where
$$
\mu = \left(
\begin{array}{c}
1 \\
-1 \\
\end{array}
\right), \quad \Sigma = \left(
\begin{array}{cc}
1 & 0.5 \\
0.5 & 1 \\
\end{array}
\right).
$$
Conditional on $X$, the substitution variable $A$ was generated from a Bernoulli
distribution such that ${\rm pr}(A=1\mid X) = \text{expit}(X^T  u ),$ where $ u = (1,1,1)^T/2$. The exposure variable $Z$ was generated following a logistic  model:
${\rm pr}(Z=1\mid X,A) = \text{expit}(\delta_1 X  u + \delta_1 A)$, where $\delta_1$ is a parameter taking values 0 or 1.  The survival type $G$ was generated  from a multinomial distribution such that ${\rm pr}(G=\textsc{LL}\mid X,A) = \text{expit} (\gamma_0 +  X^T \gamma_1  +A\gamma_2)  \text{expit}( \beta_{10}  + X^T \beta_{11} +A\beta_{12}) $, 	${\rm pr}(G=\textsc{LD}\mid X,A) = \left\{1-\text{expit} (\gamma_0 + X^T \gamma_1  +A\gamma_2) \right\}\text{expit}( \beta_{10}  + X^T \beta_{11} +A\beta_{12})$ and ${\rm pr}(G=\textsc{DD}\mid X,A)  = 1- \text{expit}( \beta_{10}  + X^T \beta_{11} +A\beta_{12})$, where $(\beta_{10}, \beta_{11}, \beta_{12}) =(4,\delta_2,\delta_2,\delta_2,2)/2$, $(\gamma_0,\gamma_1, \gamma_2) = (0, -3\delta_2,\delta_2,\delta_2,2)/2$ and $\delta_2$ is a parameter taking values 0 or 1.  
The potential outcomes $Y(z)$ were generated from the following normal
distributions: $Y(1)\mid G=\textsc{LL},X,A \sim N(X^T  u,0.5^2)$, $Y(1)\mid G=\textsc{LD},X,A \sim N(1+ X^T  u, 0.5^2)$ and $Y(0)\mid G=\textsc{LL},X,A \sim N(-1 + X^T  u, 0.5^2)$. The observed survival $S$ and observed outcome $Y$ follow  from these by the consistency assumption. Note under our settings, the treatment assignment and outcome are confounded when $\delta_1=1$, whereas
the survival and outcome processes are confounded when $\delta_2=1$.  The true value for $\Delta_{\textsc{LL}}$  is 1. 

We compared four methods for estimating $\Delta_{\textsc{LL}}$:
linear regression of $Y$ on $Z$, $X$ and $A$ among observed survivors;
the estimation method of  \cite{ding2011identifiability};  the proposed method under the exclusion restriction assumption \ref{assump.A2.x}, i.e., the estimation method using \ref{m1.1}--\ref{m2.2};   the proposed method under the no-interaction assumption \ref{assump.A2.x.alter}, i.e., the estimation method using \ref{m2.1}, \ref{m2.2} and M.7.  We differ from \cite{ding2011identifiability} in both the identification conditions and the estimation method. 
Table \ref{result} summarizes the results.  The naive regression method is biased in all settings due to selection bias. 
In the presence of confounding between $Z$ and $Y$ or $S$ and $Y$, that is, when $\delta_1 \neq 0$ or $\delta_2 \neq 0$, \cite{ding2011identifiability}'s estimator is inconsistent. However, even in the absence of confounding between $Z$ and $Y$ or $S$ and $Y$, that is, when $\delta_1=\delta_2=0$, 
\cite{ding2011identifiability}'s estimator  can be unstable, especially if the sample size is small.
Although in this case the  substitution relevance assumption of \cite{ding2011identifiability}  holds  since ${\rm pr}(G=\textsc{LL}\mid Z=1,S=1,A=1) \neq {\rm pr}(G=\textsc{LL}\mid Z=1,S=1,A=0)$, 
for some simulated samples  the estimates of ${\rm pr}(G=\textsc{LL}\mid Z=1,S=1,A=1) $ can be very close to the estimates of ${\rm pr}(G=\textsc{LL}\mid Z=1,S=1,A=0)$, leading to instability in the causal effect estimates.  {In contrast, the proposed estimators  are more stable, and have bias that decreases with sample size in all the settings considered here.}

\begin{table} 
	\begin{center}
		\caption{{{Bias times 10 and standard error times 10 (in parenthesis)} for various methods of estimating  $\Delta_{\textsc{LL}}$   under settings following \ref{assump.monotone}--\ref{assump.A3.x}. In these settings Prop-ER and Prop-NI are expected to be consistent. Here $\delta_1 = 1$ and $\delta_2 = 1$ correspond to the presence of confounding between $Z$ and $Y$, and  the presence of confounding between $S$ and $Y$, respectively.  {Results are based on 500 simulated data sets}}}
		\begin{tabular}{rccccccccccc}
\multicolumn{7}{l}{Exclusion restriction: True} \\[5pt]
			Sample size &	$\delta_1$  &  $\delta_2$  &&       \multicolumn{4}{c}	{ Estimation method} \\[5pt]
&
			&&
			&  Naive  &  DGYZ  &  Prop-ER &  Prop-NI \\[5pt]
			\quad\quad 
200 & 0  & 0   && 73(0.61)    &  3800(3300) &    7.2(7.7) &    32(1.9) \\
&    & 1   && 46(0.67)    & $-$1100(1000) &     10(1.2) &  2.9(0.91) \\
& 1  & 0   && 80(0.85)    &    160(4.7) &    $-$7.6(27) &    45(2.4) \\
&    & 1   && 40(0.83)    &     35(8.4) &     35(1.5) &    28(1.2) \\ [5pt]
1000& 0  & 0   && 73(0.27)    &    410(7.0) &   $-$4.3(1.5) &   8.3(1.1) \\
&    & 1   && 48(0.29)    &   $-$320(860) &   3.3(0.56) &  1.6(0.41) \\
& 1  & 0   && 80(0.36)    &    180(1.6) &   $-$8.3(3.5) &   9.7(1.3) \\
&    & 1   && 41(0.36)    &     79(2.5) &    11(0.97) &  9.3(0.58) \\ [5pt]
5000& 0  & 0   && 73(0.11)    &    380(2.1) & $-$0.16(0.58) &  1.7(0.48) \\
&    & 1   && 48(0.13)    &  $-$1800(810) &  0.72(0.27) & 0.37(0.21) \\
& 1  & 0   && 81(0.17)    &   180(0.65) &  $-$1.1(0.62) &  1.9(0.61) \\
&    & 1   && 41(0.16)    &     83(1.0) &   2.6(0.50) &  2.0(0.24) \\
		\end{tabular}
		\label{result}
	\end{center}
			Naive: linear regression among observed survivors; DGYZ: \cite{ding2011identifiability}'s method; Prop-ER: the proposed method assuming the exclusion restriction; Prop-NI: the proposed method assuming no interaction.
	\end{table}


\subsection{Settings with exclusion restriction violation}

We now evaluate the sensitivity of the proposed method to departures from the exclusion restriction assumption \ref{assump.A2.x}. We generated data in the same way as in Section \ref{sec:simu_1}, except that the potential outcome $Y(z)$ was generated from the following normal
distributions: $Y(1)\mid G=\textsc{LL},X,A \sim N(5+ X^T  u + A,0.5^2)$, $Y(1)\mid G=\textsc{LD},X,A \sim N(7+ X^T  u + A, 0.5^2)$ and $Y(0)\mid G=\textsc{LL},X,A \sim N(4 + X^T  u + A, 0.5^2)$.  The method of \cite{ding2011identifiability} for dealing with exclusion restriction violations is not applicable here as it requires the substitution variable $A$ to be either continuous or have at least three categories. 

The simulation results are presented in Table \ref{result_ER_vio}.  As expected, the proposed estimator assuming no interaction  is consistent for all simulation settings considered here, whereas both \cite{ding2011identifiability}'s estimator and the proposed estimator assuming the exclusion restriction are biased for estimating $\Delta_{\textsc{LL}}$. 


\begin{table} 
	\begin{center}
		\caption{{Bias times 100 and standard error times 100 (in parenthesis) for various methods of estimating  $\Delta_{\textsc{LL}}$  under settings with exclusion restriction violation. In these settings Prop-NI is expected to be consistent but Prop-ER is not. Here $\delta_1 = 1$ and $\delta_2 = 1$ correspond to the presence of confounding between $Z$ and $Y$, and  the presence of confounding between $S$ and $Y$, respectively.  Results are based on 500 simulated data sets}}
		\begin{tabular}{rccccccccccc}
\multicolumn{7}{l}{Exclusion restriction: False} \\[5pt]
			Sample size &	$\delta_1$  &  $\delta_2$  &&       \multicolumn{4}{c}	{ Estimation method} \\[5pt]
&
			&&
		&  Naive	&  DGYZ  &  Prop-ER &  Prop-NI  \\[5pt]
			\quad\quad 
200 & 0  & 0   && 73(0.61)    &  5500(4700) &   160(11)   &    32(1.9) \\
&    & 1   && 46(0.67)    & $-$1700(1500) &   16(1.7)   &  2.9(0.91) \\
& 1  & 0   && 80(0.85)    &     66(7.6) &   140(13)   &    45(2.4) \\
&    & 1   && 40(0.83)    &    $-$140(13) &   54(2.0)   &    28(1.2) \\ [5pt]
1000& 0  & 0   && 73(0.27)    &     580(10) &  160(1.9)   &   8.3(1.1) \\
&    & 1   && 48(0.29)    &  $-$540(1300) & 8.1(0.62)   &  1.6(0.41) \\
& 1  & 0   && 80(0.36)    &     94(2.4) &  170(5.3)   &   9.7(1.3) \\
&    & 1   && 41(0.36)    &    $-$62(3.8) &   31(1.6)   &  9.3(0.58) \\ [5pt]
5000& 0  & 0   && 73(0.11)    &    540(3.1) & 160(0.69)   &  1.7(0.48) \\
&    & 1   && 48(0.13)    & $-$2800(1200) & 4.7(0.27)   & 0.37(0.21) \\
& 1  & 0   && 81(0.17)    &    95(0.94) & 160(0.77)   &  1.9(0.61) \\
&    & 1   && 41(0.16)    &    $-$56(1.6) &  20(0.47)   &  2.0(0.24) \\
		\end{tabular}
		\label{result_ER_vio}
	\end{center}
		Naive: linear regression among observed survivors; DGYZ: \cite{ding2011identifiability}'s method; Prop-ER: the proposed method assuming the exclusion restriction; Prop-NI: the proposed method assuming no interaction.
\end{table}

\section{Application to a Southwest Oncology Group trial}
\label{sec:nacc}

We illustrate the advantage of the proposed  methods using data from a randomized phase III trial to compare docetaxel plus
estramustine with mitoxantrone plus prednisone in men with metastatic, hormone independent
prostate cancer \citep{petrylak2004docetaxel}.  The  data set we use, which was created by \cite{ding2011identifiability}, contains observations on 487 men aged from 47 to 88. Of these subjects, 258 were randomly assigned to receive docetaxel plus
estramustine  and 229 were randomly assigned to receive mitoxantrone plus prednisone.  In our analysis, we are interested in comparing these two treatments in terms of the health related quality of life  one year after receiving the treatment. 

A naive analysis shows that among patients who survived to one year after receiving the assigned treatment, the quality of life for those assigned to the \tx group is higher by 2.46  units  compared to those assigned to the \ctrl\ group; the 95\% confidence interval is [$-$3.31,8.24]. However, this estimate is not causally interpretable as subjects who would survive if assigned to \tx are potentially different from subjects who would survive if assigned to \ctrl. Moreover, as reported by \cite{petrylak2004docetaxel}, \tx  is beneficial for the overall survival compared to \ctrl.  The direct comparison among observed survivors is hence also subject to selection bias. Instead, we apply  \cite{ding2011identifiability}'s and our proposed methods to deal with truncation due to death. To account for possible common causes of the survival and outcome processes, we adjust for the following variables in the proposed methods: age, race, type of prognosis, bone pain  and performance status.  Following \cite{ding2011identifiability}, we use the baseline quality of life as the substitution variable, and the change in quality of life in the one-year period as the outcome.

We first analyze this data set under the monotonicity assumption.  The point estimates and bootstrap standard errors are displayed in Table \ref{table:ding}.  The results in \cite{ding2011identifiability} suggest that \tx had a significant causal effect on the quality of life among those who would survive one year after receipt of treatment regardless of which treatment group they were assigned to. In contrast, after accounting for the baseline covariate information that might simultaneously impact the potential survival and  the potential quality of life, we were not able to reach such a conclusion. Both of the proposed methods yield a point estimate that is much closer to 0, and their 95\% confidence intervals cover 0.  These results show that, even in the setting of a randomized trial, adjusting for baseline covariates can lead to different estimates for the survivor average causal effect;  this should not be surprising since baseline variables may confound $S$ and $Y$.

\begin{table}
	\begin{center}
		\caption{Survivor average causal effect of \tx on health related quality of life}
		\label{table:ding}
		\begin{tabular}{rccccc}
			Estimation method & Point estimate &  \text{Bootstrapped SE} & 2.5\%  &  50\% &  97.5\% \\[5pt]
			 DGYZ & 7.01 & 3.09 & 1.81 & 6.56 & 13.64 \\ 
			 Prop-ER & 3.06 & 11.79 & $-$15.15 & 3.29 & 22.60 \\ 
			Prop-NI & 2.73 & 3.82 & $-$3.97 & 2.95 & 10.83 \\ 				
		\end{tabular}
	\end{center}
 DGYZ: results adapted from \cite{ding2011identifiability}; Prop-ER: results estimated using the proposed method assuming the exclusion restriction; Prop-NI: results estimated using the proposed method assuming no interaction.
\end{table}

\begin{figure}[!htbp]
	\centering
	\includegraphics[width=.8\linewidth]{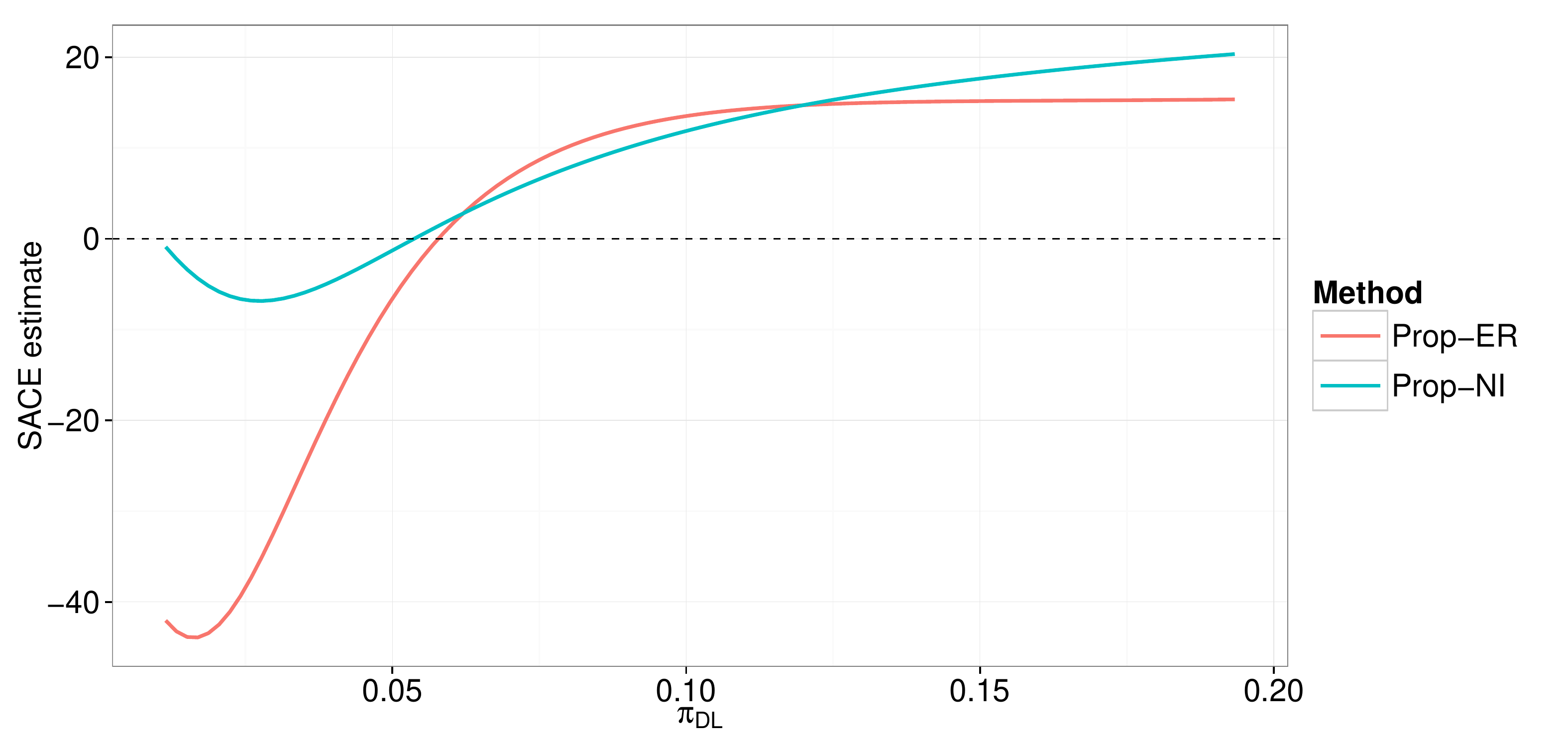}	
	\caption{Sensitivity analysis for estimating the survivor average causal effect in the Southwest Oncology Group dataset. Solid line: the proposed method assuming the exclusion restriction; dashed line: the proposed method assuming no interaction.  		}
	\label{fig:ding_sens}
\end{figure}

The proposed estimator Prop-ER, that assumes  the exclusion restriction and the monotonicity assumption, is very unstable.
In particular, the point estimate for the model parameter $\gamma$ in \ref{m2.2} is large in magnitude, so that some fitted values of $\theta_{0/1}$ are close to the boundary. This indicates a monotonicity violation. In fact, previous analyses also suggest that the monotonicity assumption might be problematic for this data set \citep{ding2011identifiability} and as we discussed earlier, for randomized trials, the monotonicity assumption is not plausible {a priori}. For these reasons, 
although the overall one year survival rate in the \tx group (49.6\%) is higher than that in the \ctrl\ group (38.9\%), a sensitivity analysis of the monotonicity assumption is warranted. 
In our analysis the sensitivity parameter is $\pi_{\textsc{DL}}$, which has a one-to-one correspondence with the sensitivity parameter $\rho$ but has the interpretation that it is the overall fraction of patients who would die one year after receiving \tx but would survive one year after receiving \ctrl. 
Under our modeling assumptions \ref{m2.1} and M.6,  the range for $\pi_{\textsc{DL}}$  is estimated to be $[0.01,0.19]$, with the lower and upper limits corresponding to $\rho = 1$ and $0$, respectively. As the lower limit is greater than 0, the monotonicity assumption \ref{assump.monotone} is not compatible with these modeling assumptions. 
Figure \ref{fig:ding_sens} summarizes the results from our sensitivity analysis. 
As both of the sensitivity lines cross 0, without prior knowledge on the possible values for $\pi_{\textsc{DL}}$, we cannot reach any definitive answer about the causal effect of \tx versus \ctrl\ on the quality of life one year after receiving the treatment. 
Furthermore, the two curves from the proposed methods agree approximately  unless $\pi_{\textsc{DL}}$ is smaller than 0.05. We also note that  under a different set of assumptions, \cite{ding2011identifiability} obtained a value of $\pi_{\textsc{DL}}$ that is much greater than our upper bound. Under our modeling assumptions \ref{m2.1} and M.6, such a value is incompatible with the stochastic monotonicity assumption \ref{assump:stochastic_monotone}, and would imply a negative  correlation between the potential survival under docetaxel plus
estramustine and the potential survival under mitoxantrone plus prednisone. 

\section{Discussion}
\label{sec:discussion}

In this article, we considered the identification and estimation of the survivor average causal effect. Compared with previous works that attempt to identify the survivor average causal effect using an exogenous or instrumental variable, our method adjusts for baseline covariates $X$, which may be common causes between any pair of treatment, survival and outcome.
 This role for baseline covariates is unique to our identification framework for the survivor average causal effect. In contrast to the standard estimation problem of the average causal effect, even in randomized studies, the inclusion of baseline covariates is crucial for obtaining unbiased estimates of the survivor average causal effect,  since it is likely there is confounding between survival and the post-survival outcome. 

In our estimation approach, instead of imposing modeling assumptions on the observed data, we use parametric models for the potential outcomes.  This not only simplifies the estimation procedure, but also makes it easier to ensure compatibility between our identification and modeling assumptions.  We also propose alternative methods to relax the monotonicity assumption and the exclusion restriction.  Formal inference under our alternative assumption \ref{assump:stochastic_monotone} requires a sensitivity analysis framework that accounts for uncertainty due to both  non-identifiability and sampling variability, such as those based on ignorance and uncertainty regions described by \cite{vansteelandt2006ignorance} and others. To focus on the main ideas here we only present point estimates under \ref{assump:stochastic_monotone} for fixed values of $\rho$.

In the current work we have only considered the binary exposure  case.  In practical medical studies, the exposure variable may have multiple levels. For example,  the smoking variable may be coded as  never smokers, past smokers and current smokers. In this case, it would be interesting to generalize  the proposed methods  to deal with ordinal exposure status.

\section*{Acknowledgements}

The authors thank the referees,  associate editor, editor and members of the National Alzheimer's Coordinating Center  for helpful comments. The authors also thank Peng Ding, Peter Gilbert, James Robins and Eric Tchetgen Tchetgen for valuable discussions.  This work was initiated when the first author was a graduate student at the University of Washington. Zhou was supported in part by the US Department of Veterans Affairs. Richardson was supported by the National Institutes of
Health. 

%


\section*{Supplementary material}

Supplementary material available at \emph{Biometrika} online includes  proofs of theorems, propositions and alternative causal diagrams that are compatible with the structural equations \eqref{eqn:npsem}, but allow for certain dependencies between error terms.

\section*{Appendix}
\renewcommand{\thesubsection}{\Alph{subsection}}

\subsection{Parameterizations under alternative identification assumptions}
\label{appendix:models_alter}

To estimate $\Delta_{\textsc{LL}}$  under the assumptions of Theorem \ref{thm:identification_alter1}, we assume \ref{m1.2}, \ref{m2.1} and the following model:
\begin{enumerate}
	\item[M.5] \label{m3.1} 
	$E\{Y(0)\mid Z=0,X,G,A\}$ is known up to a finite-dimensional parameter $\alpha_3$; that is, $E\{Y(0)\mid Z=0,X,G,A\} = m_{3}(X,G; \alpha_3)$, where $G$ takes values in $\{\textsc{LL},\textsc{DL}\}$, $m_{3}(\cdot,\cdot; \alpha_3)$ is a known function and $\alpha_3$ is an unknown parameter.  Note that due to condition (i) of  Theorem \ref{thm:identification_alter1}, $m_{3}(\cdot,\cdot; \alpha_3)$ does not depend on the value of $A$. 
	Specifically, for the simulations and data example we code $\textsc{LL}$ to be 1, $\textsc{DL}$ to be 0, and consider
	\begin{equation*}
	\label{eqn:m3.1}
	m_{3}(X,G; \alpha_3) = \alpha_{30} +  X^T \alpha_{31} +  G \alpha_{33}.
	\end{equation*}

	\item [M.6]\label{m3.2}  
	${\rm pr}\{S(0)=1\mid X,A\}$ is known up to a finite-dimensional parameter $\beta_0$; that is ${\rm pr}\{S(0)=1\mid X,A\} = \theta_0(X,A;\beta_0)$, where $\theta_0(\cdot,\cdot;\beta_0)$ is a known function and $\beta_0$ is an unknown parameter.
	Specifically, for the  data example we consider
	\begin{equation*}
	\label{eqn:m3.2}
	\theta_0(X,A;\beta_0) = \text{expit}(\beta_{00}+  X^T \beta_{01} + A \beta_{02}).
	\end{equation*}
\end{enumerate}
Similarly, to estimate  $\Delta_{\textsc{LL}}$  under the assumptions of Theorem \ref{thm:identification_alter},  we assume \ref{m2.1}, \ref{m2.2} and the following model:
\begin{enumerate}
	\item[M.7] \label{m4.1} $E(Y\mid X,A,G,Z)$ is known up to a finite-dimensional parameter $\alpha_4$; that is, $E[Y\mid X,A,G,Z] = m_{4}(X,A,G,Z; \alpha_4)$, where $G$ takes values in $\{\textsc{LL},\textsc{LD}\}$, $m_{4}(\cdot,\cdot,\cdot,\cdot; \alpha_4)$ is a known function and $\alpha_4$ is an unknown parameter.  Note that due to \ref{assump.A2.x.alter}, $m_{4}(X,A,G,Z; \alpha_4)$ should not contain interaction terms within the pairs $(A,Z)$ or $(A,G)$.
	Specifically, for the simulations and data example we code $\textsc{LL}$ to be 1, $\textsc{LD}$ to be 0 and consider
	\begin{equation*}
	\label{eqn:m4.1}
	m_{4}(X,A,G,Z; \alpha_4) = \alpha_{40} +  X^T \alpha_{41} + A \alpha_{42} + G\alpha_{43} + Z \alpha_{44}.
	\end{equation*}
\end{enumerate}
Finally, if one wishes to relax both the monotonicity assumption \ref{assump.monotone} and the exclusion restriction  \ref{assump.A2.x}, one may assume \ref{m2.1}, M.6, and the following model:
\begin{enumerate}
	\item[M.8] \label{m5.1} $E(Y\mid X,A,G,Z)$ is known up to a finite-dimensional parameter $\alpha_5$; that is, $E(Y\mid X,A,G,Z) = m_{5}(X,A,G,Z; \alpha_5)$, where  $G$ takes values in $\{\textsc{LL},\textsc{LD},\textsc{DL}\}$, $m_{5}(\cdot,\cdot,\cdot,\cdot; \alpha_5)$ is a known function and $\alpha_5$ is an unknown parameter.   Note that due to \ref{assump.A2.x.alter}, $m_{5}(X,A,G,Z; \alpha_5)$ should not contain interaction terms within the pairs $(A,Z)$ or $(A,G)$. Specifically, for the  data example we code $\textsc{LL}$ to be 1, $\textsc{DL}$ and $\textsc{LD}$ to be 0 and consider
	\begin{equation*}
	\label{eqn:m5.1}
	E(Y\mid X,A,G,Z) = \alpha_{50} +  X^T \alpha_{51} + A \alpha_{52} + ZG\alpha_{53} + Z \alpha_{54} + (1-Z)G \alpha_{55}.
	\end{equation*}
\end{enumerate}

\bibliographystyle{biometrika}
\bibliography{causal}

\clearpage
\thispagestyle{empty}
\begin{center}
	{\Large Supplementary Materials for ``Identification and Estimation of Causal Effects with Outcomes Truncated by Death''}
		\bigskip
	
	{\Large  Linbo Wang, Xiao-Hua Zhou \emph{and} Thomas Richardson } $\ $
	
\end{center}
\setcounter{equation}{0}
\setcounter{figure}{0}
\setcounter{table}{0}
\setcounter{page}{1}
\makeatletter
\renewcommand{\theequation}{S\arabic{equation}}
\renewcommand{\thefigure}{S\arabic{figure}}
\setcounter{section}{0}
%

\begin{abstract}
	In this Supplementary Material we provide  proofs of  theorems and propositions 
	in the paper, and alternative causal diagrams that are compatible with the structural equations (4) in the paper,
	but allow for certain dependencies between error terms.
	Some additional equations and figures are created within this document, which
	we label as equation (S1),  Figure S1 and so forth, to distinguish them 
	from those in the main text.
\end{abstract}

\section{Proof of Theorems 1--3}

By definition, $\Delta_{\textsc{LL}} = E\{Y(1)\mid G=\textsc{LL}\} - E\{Y(0)\mid G=\textsc{LL}\}$, where  $E\{Y(z)\mid G=\textsc{LL}\}$ is given in (1). It then suffices to identify $\pi_{\textsc{LL}\mid W}$ and $\mu_{z,\textsc{LL},W}$ for identification of $\Delta_{\textsc{LL}}$.

\emph{Identification of $\pi_{\textsc{LL}\mid W}$}: under the assumptions of Theorem 1 or 3,  $\pi_{\textsc{LL}\mid W}$ can be identified from (2), whereas under the assumptions of Theorem 2, $\pi_{\textsc{LL}\mid W}$  can be 
identified from  the following equations:
\begin{flalign*}
{\rm pr}\{S(1)=1\mid W\} &= {\rm pr}(S=1\mid Z=1,W) = \pi_{\textsc{LL}\mid W} + \pi_{\textsc{LD}\mid W}; \\
{\rm pr}\{S(0)=1\mid W\} &= {\rm pr}(S=1\mid Z=0,W) = \pi_{\textsc{LL}\mid W} + \pi_{\textsc{DL}\mid W}; \\
\pi_{\textsc{LL}\mid W} &= {\rm pr}\{S(0)=1\mid W\} {\rm pr}\{S(1)=1\mid S(0)=1,W\}; \\[4pt]
{\rm pr}\{S(1)=1\mid S(0)=1,W\} &= {\rm pr}\{S(1)=1\mid W\}\\
&\quad\quad\quad + \rho(W) \left(\min\left[1, \dfrac{{\rm pr}\{S(1)=1 \mid W\}}{{\rm pr}\{S(0)=1 \mid W\}}\right] - {\rm pr}\{S(1)=1\mid W\} \right),
\end{flalign*}
where due to A.2, ${\rm pr}\{S(z)=1\mid W\} = {\rm pr}(S=1\mid Z=z, W)$ and thus is identifiable for $z=0,1$.
Note that under these assumptions, $\pi_{g\mid W}$ is also identifiable with $g\in \{\textsc{LD}, \textsc{DL}, \textsc{DD}\}$.

\emph{Identification of $\mu_{z,\textsc{LL},W}$}: under the assumptions of Theorem 1, $\mu_{0,\textsc{LL},W}$ can be identified from (3).  
First recall that $W=(A,X)$.
As described above, $p_{g\mid z,w, s\!=\!1} = {\rm pr}(G=g\mid X=x,A=a)/{\rm pr}(S=1\mid Z=z,X=x,A=a) $ is identifiable from the observed data. Note that by consistency, $0=p_{\textsc{DL}\mid 1,x,a,s\!=\!1} = p_{\textsc{LD}\mid 0,x,a,s\!=\!1}=p_{\textsc{DD}\mid z,x,a,s\!=\!1}$, $z\in\{0,1\}$.
Furthermore, A.5 implies that for all $x$, there exist $a_0\neq a_1$ such that $p_{\textsc{LL}\mid 1,x,a_1,s\!=\!1}\neq p_{\textsc{LL}\mid 1,x,a_0,s\!=\!1}$. Due to A.4, for $g=\textsc{LL},\textsc{LD},$
\begin{equation*}
\mu_{1,g,w} = E(Y\mid Z=1,G=g,X=x,A=a) = E(Y\mid Z=1,G=g,X=x) \equiv \mu_{1,g,x}. 
\end{equation*}
Furthermore, $\mu_{1,\textsc{LL},x}$ can be identified from
\begin{flalign*}
E(Y\mid Z=1,S=1,X=x,A=a_1) &= p_{\textsc{LL}\mid 1,x,a_1,s\!=\!1} \mu_{1,\textsc{LL},x} + (1-p_{\textsc{LL}\mid 1,x,a_1,s\!=\!1}) \mu_{1,\textsc{LD},x},  \\
E(Y\mid Z=1,S=1,X=x,A=a_0) &= p_{\textsc{LL}\mid 1,x,a_0,s\!=\!1} \mu_{1,\textsc{LL},x} + (1-p_{\textsc{LL}\mid 1,x,a_0,s\!=\!1})\mu_{1,\textsc{LD},x}. \numberthis \label{eqn:identification}
\end{flalign*}

Under the assumptions of Theorem 2, we can obtain similarly that for $z\in\{0,1\}$, $\mu_{z,\textsc{LL},w} = \mu_{z,\textsc{LL},x}$, where $\mu_{1,\textsc{LL},x}$ is identified by \eqref{eqn:identification}, and $\mu_{0,\textsc{LL},x}$ can be identified from
\begin{flalign*}
E(Y\mid Z=0,S=1,X=x,A=a_1^\prime) &= p_{\textsc{LL}\mid 0,x,a_1^\prime,s\!=\!1} \mu_{0,\textsc{LL},x} + (1-p_{\textsc{LL}\mid 0,x,a_1^\prime,s\!=\!1}) \mu_{0,\textsc{DL},x},  \\
E(Y\mid Z=0,S=1,X=x,A=a_0^\prime) &= p_{\textsc{LL}\mid 0,x,a_0^\prime,s\!=\!1} \mu_{0,\textsc{LL},x} + (1-p_{\textsc{LL}\mid 0,x,a_0^\prime,s\!=\!1})\mu_{0,\textsc{DL},x},
\end{flalign*}
where $a_1^\prime$ and $a_0^\prime$ are two distinct values of $A$ such that $p_{\textsc{LL}\mid 0,x,a_1^\prime,s\!=\!1}\neq p_{\textsc{LL}\mid 0,x,a_0^\prime,s\!=\!1}$.

Similarly, under the assumptions of Theorem 3, $\mu_{0,\textsc{LL},W}$ is identified by (3) and  $\mu_{1,\textsc{LL},x,a_1}$  can be identified from the following equations:
\begin{flalign*}
E(Y\mid Z=1,S=1,X=x,A=a_1) &= p_{\textsc{LL}\mid 1, x,a_1,s\!=\!1} \mu_{1,\textsc{LL},x,a_1} + (1-p_{ \textsc{LL}\mid 1, x,a_1,s\!=\!1}) \mu_{1, \textsc{LD},x,a_1};  \\
E(Y\mid Z=1,S=1,X=x,A=a_0) &= p_{\textsc{LL}\mid 1, x,a_0,s\!=\!1} \mu_{1, \textsc{LL},x,a_0} + (1-p_{\textsc{LL}\mid 1, x, a_0,s\!=\!1})\mu_{1, \textsc{LD},x,a_0}; \\
\mu_{1, \textsc{LL},x,a_1} - \mu_{1, \textsc{LL},x,a_0} &= \mu_{1, \textsc{LD},x,a_1} - \mu_{1, \textsc{LD},x,a_0}; \\
\mu_{1, \textsc{LL},x,a_1} - \mu_{1, \textsc{LL},x,a_0} &= \mu_{0, \textsc{LL},x,a_1} - \mu_{0, \textsc{LL},x,a_0}; \\
E(Y\mid Z=0,S=1,X=x,A=a_1) &= \mu_{0, \textsc{LL},x,a_1}; \\
E(Y\mid Z=0,S=1,X=x,A=a_0) &= \mu_{0, \textsc{LL},x,a_0},
\end{flalign*}
where $\mu_{z,g,x,a} \equiv E(Y\mid Z=z,G=g,X=x,A=a)$.

For any $a\in \mathcal{A}\mathbin{\backslash} \{a_1\}$, $\mu_{1,\textsc{LL},x,a}$ can be identified via the following equations:
\begin{flalign*}
\mu_{1, \textsc{LL},x,a} - \mu_{1, \textsc{LL},x,a_1} &= \mu_{0, \textsc{LL},x,a} - \mu_{0, \textsc{LL},x,a_1}; \\
E(Y\mid Z=0,S=1,X=x,A=a) &= \mu_{0, \textsc{LL},x,a}; \\
E(Y\mid Z=0,S=1,X=x,A=a_1) &= \mu_{0, \textsc{LL},x,a_1}.
\end{flalign*}

\section{Proof that the nonparametric structural equation model with independent errors  in Section 3.3 implies A.2--A.4}
\label{appendix:npsem}

Under (4),  $\epsilon_Z \ind (\epsilon_X, \epsilon_A, \epsilon_S, \epsilon_Y)$ implies that $Z \ind (\epsilon_S, \epsilon_Y) \mid X, A$, which then implies A.2 and A.3; recall here that the error terms can be interpreted as the set of one-step-ahead counterfactuals.
To see that A.4 holds under the structural equations (4), note the following:
\begin{flalign*}
{\rm pr}\{Y \mid  A=a_1, Z=1, X=x, S(1) = 1, S(0)\}&= {\rm pr}\{Y \mid  A=a_1, Z=1, X=x, S = 1, S(0)\}  \\
&= {\rm pr}(Y \mid  A=a_1, Z=1, X=x, S=1) \\
&= {\rm pr}(Y \mid  A=a_0, Z=1, X=x, S=1)  \\
&= {\rm pr}\{Y \mid  A=a_0, Z=1, X=x, S = 1, S(0)\}  \\
&= {\rm pr}\{Y \mid  A=a_0, Z=1, X=x, S(1) = 1, S(0)\},
\end{flalign*}
where the first and last lines follow from consistency; the second and fourth follow from
$Y \ind S(0)\mid A, Z=1, X, S=1$, which holds as $Y$ is d-separated from $\epsilon_S$ given $X,A,Z,S$; the third line follows since $Y$ is d-separated from $A$ conditioning on $X, Z, S$.

As A.4 holds trivially for the subgroup with $S(1) = 0$, we complete the proof. \qed

\section{Proof of Proposition 1}
\label{appendix:proof_prop_constriants}

Constraint (8) is standard.   To prove (9), note that under A.4, \eqref{eqn:identification} holds for all $a$, i.e.
\begin{equation}
\label{eqn:13}
E(Y\mid Z=1,S=1,X=x,A=a) = p_{\textsc{LL}\mid 1,x,a,s\!=\!1} \mu_{1,\textsc{LL},x} + (1-p_{\textsc{LL}\mid 1,x,a,s\!=\!1}) \mu_{1,\textsc{LD},x}.
\end{equation}
It follows that
$$
| 	E(Y\mid Z=1,S=1,X=x,A=a) | \leq \max(|\mu_{1,\textsc{LL},x}|, |\mu_{1,\textsc{LD},x}|).
$$
Hence for all $x$, $	E(Y\mid Z=1,S=1,X=x,A=a)$ is bounded as a function of $a$. On the other hand, suppose that  for all $x$, $	E(Y\mid Z=1,S=1,X=x,A=a)$ is bounded as a function of $a$. Let
$$
\bar{f}(x) = \sup\limits_a E(Y\mid Z=1,S=1,X=x,A=a), \text{\; and}\; \ubar{f}(x) = \inf\limits_a E(Y\mid Z=1,S=1,X=x,A=a).
$$
Then \eqref{eqn:13} holds with
$$
p_{\textsc{LL}\mid 1,x,a,s\!=\!1} = \dfrac{\bar{f}(x) - E(Y\mid Z=1,S=1,X=x,A=a)}{\bar{f}(x) - \ubar{f}(x)}, \; \mu_{1,\textsc{LL},x} = \ubar{f}(x) \text{\; and \;}
\mu_{1,\textsc{LD},x} = \bar{f}(x).
$$
Hence (9) summarizes all the constraints  on the observed data law derived from \eqref{eqn:13}.  The proof of (11) is similar and hence omitted.

Constraint (10) follows immediately from A.5 by noting that 
\begin{flalign*}
\dfrac{{\rm pr}(S=1\mid Z=0,X=x,A=a)}{{\rm pr}(S=1\mid Z=1,X=x,A=a)} &= 		\dfrac{{\rm pr}(G=\textsc{LL}\mid X=x,A=a)}{{\rm pr}(S=1\mid Z=1,X=x,A=a)}\\ &= {\rm pr}(G=\textsc{LL}\mid Z=1, S=1, X=x, A=a).
\end{flalign*}

To show (12), note that A.7 implies that for all $a_1,a_0$,
\begin{flalign*}
\mu_{1,\textsc{LL},x,a_1}-\mu_{0,\textsc{LL},x,a_1} &= 	\mu_{1,\textsc{LL},x,a_0}-\mu_{0,\textsc{LL},x,a_0}, \\
\mu_{1,\textsc{LD},x,a_1}-\mu_{0,\textsc{LL},x,a_1} &= 	\mu_{1,\textsc{LD},x,a_0}-\mu_{0,\textsc{LL},x,a_0}.
\end{flalign*}
It follows that $\mu_{1,\textsc{LL},w} - \mu_{0,\textsc{LL},w} = \mu_{1,\textsc{LL},x} - \mu_{0,\textsc{LL},x}$ and $\mu_{1,\textsc{LD},w} - \mu_{0,\textsc{LL},w} = \mu_{1,\textsc{LD},x} - \mu_{0,\textsc{LL},x}$. We then have 
\begin{flalign*}
&\quad	E(Y\mid Z=1,S=1,X=x,A=a) - E(Y\mid Z=0,S=1,X=x,A=a) \\ 
&= p_{\textsc{LL}\mid 1,x,a,s\!=\!1} \mu_{1,\textsc{LL},x,a} + p_{\textsc{LD}\mid 1,x,a,s\!=\!1} \mu_{1,\textsc{LD},x,a} - \mu_{0,\textsc{LL},x,a} \\
&= p_{\textsc{LL}\mid 1,x,a,s\!=\!1} (\mu_{1,\textsc{LL},x,a} - \mu_{0,\textsc{LL},x,a}) + p_{\textsc{LD}\mid 1,x,a,s\!=\!1}(\mu_{1,\textsc{LD},x,a} - \mu_{0,\textsc{LL},x,a}) \\
&= p_{\textsc{LL}\mid 1,x,a,s\!=\!1} (\mu_{1,\textsc{LL},x} - \mu_{0,\textsc{LL},x}) + p_{\textsc{LD}\mid 1,x,a,s\!=\!1}(\mu_{1,\textsc{LD},x} - \mu_{0,\textsc{LL},x})
\end{flalign*}
so that 
\begin{flalign*}
&\quad |E(Y\mid Z=1,S=1,X=x,A=a) - E(Y\mid Z=0,S=1,X=x,A=a)|  \\ &\leq \max(|\mu_{1,\textsc{LL},x} - \mu_{0,\textsc{LL},x}|,|\mu_{1,\textsc{LD},x} - \mu_{0,\textsc{LL},x}|).
\end{flalign*}
The rest of the proof is similar to that for (9) and is hence omitted.

\section{Alternative causal diagrams}
\label{appendix:npsem_alter}

Figure \ref{fig:npsem_alter} shows some more complicated causal diagrams that imply assumptions A.2--A.4 but allow for dependence of errors in the structural equations (4) representing unmeasured confounding  between nodes in Fig. 1. Here unmeasured confounding is denoted by a bi-directed edge between observed nodes. In general, using  graphical terminology \citep{richardson2003markov},  the mixed graphs, such as those in Fig. \ref{fig:npsem_alter},  imply A.2--A.4  so long as the pairs $(Z,S)$, $(Z,Y)$, $(A,Y)$ and $(S,Y)$ are not in the same district, where a district is a connected component of the graph obtained by removing all edges that are not bi-directed.  Specifically, A.2 requires that  $Z$ and $S$ cannot be in the same district, A.3 requires that neither $Z$ and $Y$ nor 
$S$ and $Y$ can be in the same district, and A.4 requires additionally that  $A$ and $Y$ cannot be in the same district. 

\begin{figure}[!htbp] 
	\centering
	\scalebox{0.8}{
		\begin{tikzpicture}[->,>=stealth',shorten >=1pt,auto,node distance=2.5cm, shape=ellipse,
		semithick, scale=0.3]
		pre/.style={->,>=stealth,semithick,blue,line width = 1pt}]
		\tikzstyle{every state}=[fill=none,draw=black,text=black, shape=ellipse]
		\node[state] (X)                    {$X$};
		\node[state] (A) [right of=X] {$A$};
		\node[state]         (Z) [right of=A] {$Z$};
		\node[state](S) [right of = Z]{$S$};
		\node[state](Y)[right of = S]{$Y$};
		\path	(X) edge node {} (A)
		(A) edge node {} (Z)
		(Z) edge node {} (S)
		(S) edge node  {} (Y);
		\draw[->] (X) to[out=20,in=160,looseness=1.1] (Z);
		\draw[->] (X) to[out=40,in=140,looseness=0.5] (S);
		\draw[->] (X) to[out=60,in=120,looseness=0.75] (Y);
		\draw[->] (A) to[out=20,in=160](S);
		\draw[->] (Z) to[out=20,in=160] (Y);
		\draw[<->] (X) to[out=-20,in=-160] (A);
		\draw[<->] (A) to[out=-20,in=-160] (Z);
		\draw[<->] (X) to[out=-40,in=-140] (Z);
		\end{tikzpicture}	
	}
	\begin{center}
		(i)
	\end{center}
	\scalebox{0.8}{
		\begin{tikzpicture}[->,>=stealth',shorten >=1pt,auto,node distance=2.5cm, shape=ellipse,
		semithick, scale=0.3]
		pre/.style={->,>=stealth,semithick,blue,line width = 1pt}]
		\tikzstyle{every state}=[fill=none,draw=black,text=black, shape=ellipse]
		\node[state] (X)                    {$X$};
		\node[state] (A) [right of=X] {$A$};
		\node[state]         (Z) [right of=A] {$Z$};
		\node[state](S) [right of = Z]{$S$};
		\node[state](Y)[right of = S]{$Y$};
		\path	(X) edge node {} (A)
		(A) edge node {} (Z)
		(Z) edge node {} (S)
		(S) edge node  {} (Y);
		\draw[->] (X) to[out=20,in=160,looseness=1.1] (Z);
		\draw[->] (X) to[out=40,in=140,looseness=0.5] (S);
		\draw[->] (X) to[out=60,in=120,looseness=0.75] (Y);
		\draw[->] (A) to[out=20,in=160] (S);
		\draw[->] (Z) to[out=20,in=160] (Y);
		\draw[<->] (X) to[out=-20,in=-160] (A);
		\draw[<->] (X)  to[out=-40,in=-160] (S);
		\end{tikzpicture}	
	}
	\begin{center}
		(ii)
	\end{center}
	\scalebox{0.8}{
		\begin{tikzpicture}[->,>=stealth',shorten >=1pt,auto,node distance=2.5cm, shape=ellipse,
		semithick, scale=0.3]
		pre/.style={->,>=stealth,semithick,blue,line width = 1pt}]
		\tikzstyle{every state}=[fill=none,draw=black,text=black, shape=ellipse]
		\node[state] (X)                    {$X$};
		\node[state] (A) [right of=X] {$A$};
		\node[state]         (Z) [right of=A] {$Z$};
		\node[state](S) [right of = Z]{$S$};
		\node[state](Y)[right of = S]{$Y$};
		\path	(X) edge node {} (A)
		(A) edge node {} (Z)
		(Z) edge node {} (S)
		(S) edge node  {} (Y);
		\draw[->] (X) to[out=20,in=160,looseness=1.1] (Z);
		\draw[->] (X) to[out=40,in=140,looseness=0.5] (S);
		\draw[->] (X) to[out=60,in=120,looseness=0.75] (Y);
		\draw[->] (A) to[out=20,in=160] (S);
		\draw[->] (Z) to[out=20,in=160] (Y);
		\draw[<->, bend left] (Z) edge (A);
		\draw[<->, bend right] (X) edge (Y);
		\end{tikzpicture}	
	}
	\begin{center}
		(iii)
	\end{center}
	\caption{More complicated causal diagrams that imply A.2--A.4. The districts in these graphs are: (i) $\{X,A,Z\}, \{S\}$ and $\{Y\}$; (ii) $\{X,A,S\}, \{Z\}$ and $\{Y\}$; (iii) $\{X,Y\}, \{A,Z\}$ and $\{S\}$.}
	\label{fig:npsem_alter}
\end{figure}

\end{document}